\documentclass{aa}
\usepackage{epsf}
\usepackage{amssymb}
\usepackage{amsmath}
\def\longversion#1{}

\def\erg{\hbox{erg}}
\def\gram{\hbox{g}}
\def\cm{\hbox{cm}}
\def\sec{\hbox{s}}

\def\xiadv{\xi}
\def\myxi{\chi}
\def\mfpeta{\eta}
\def\mnras{MNRAS}
\def\apj{ApJ}
\def\aap{A\&A}
\begin{document}
\thesaurus{(02.01.2,02.02.1)}
%\title{A semi--analytical model of disk evaporation}
\title{A semi--analytical model of disk evaporation by thermal conduction}
\titlerunning{}
\author{C.P. Dullemond}
\institute{Leiden Observatory, P.O. Box 9513, 2300 RA Leiden, The Netherlands}
\date{DRAFT, \today}
\maketitle

\begin{abstract}  
The conditions for disk evaporation by electron thermal conduction are
examined, using a simplified semi--analytical 1-D model. The model is
based on the mechanism proposed by Meyer \& Meyer--Hofmeister
(\cite{meyermeyhof:1994}) in which an advection dominated accretion
flow evaporates the top layers from the underlying disk by thermal
conduction. The evaporation rate is calculated as a function of the
density of the advective flow, and an analysis is made of the time
scales and length scales of the dynamics of the advective flow. It is
shown that evaporation can only completely destroy the disk if the
conductive length scale is of the order of the radius. This implies
that radial conduction is an essential factor in the evaporation
process. The heat required for evaporation is in fact produced at
small radii and transported radially towards the evaporation region.
\end{abstract}


\section{Introduction}
Ever since their theoretical rediscovery, advection dominated
accretion flows (ADAFs, Abramowicz et al.~\cite{abrchen:1995};
Narayan \& Yi \cite{narayanyi:1994},\cite{narayanyi:1995}; Ichimaru
\cite{ichimaru:1977}) have been widely regarded as the most likely
source of Comptonized X-ray radiation observed from many X-ray
binaries. On the basis of observational evidence (Lasota
\cite{lasota:1996a}, Narayan et al.~\cite{narbarclint:1997}, Hameury
et al.~\cite{hamlasclnar:1997} and references therein) it is believed
that ADAFs form the inner part of the accretion disk system, while the
outer part is formed by a standard Shakura--Sunyaev disk (SSD, Shakura
\& Sunyaev \cite{shaksuny:1973}).  Despite the fact that such a
bimodal disk geometry has already been proposed a long time ago
(Thorne \& Price, \cite{thorneprice:1975}; Shapiro, et
al.~\cite{sle:1976}, henceforth SLE), no satisfactory theoretical
explanation for these disk transitions has so far been found.

It seems plausible that disk surface evaporation is responsible for
the transition. By some mechanism, the uppermost layers of the disk
are heated up faster than radiative cooling can cool them down.  The
resulting hot `vapor' forms a corona on top of the disk. Part of this
vapor then accretes towards the central object, while another part
moves outwards via a transsonic wind or breeze (Meyer \&
Meyer--Hofmeister \cite{meyermeyhof:1994}, 
Liu et al.~\cite{liumeymey:1997}, 
Dullemond \& Turolla \cite{dulturigum:1997}).
At a certain radius $R_{evap}$ the entire disk has been evaporated,
and the `corona' therefore becomes a true ADAF within this evaporation
radius.

It remains uncertain what drives the surface evaporation.  Several
mechanisms have been proposed.  When no corona exists beforehand, one
can show that the upper layers of the SSD are unstable with respect to
thermal perturbations (Shaviv \& Wehrse \cite{shavivwehrse:1986};
Hubeny \cite{hubeny:1989}; Tsch\"ape \& Kley \cite{tschapekley:1993}).
These layers will heat up in a run-away fashion, thereby causing the
evaporation of the upper layers of the disk. Also, acoustic effects
(Icke \cite{icke:1976}) and magneto-hydrodynamical fluctuations and
instabilities may produce a corona, in a way similar to the formation
of the solar corona (Galeev et al.~\cite{galeev:1979}; Tout \& Pringle
\cite{toutpring:1992}).

The most promising mechanism for disk evaporation is electron thermal
conduction (Meyer \& Meyer--Hofmeister \cite{meyermeyhof:1994},
henceforth MMH). A pre--existing hot `corona' injects energy into an
extremely thin layer on top of the underlying disk, turning this
material into new hot coronal plasma. This mechanism is studied in
this paper.

In their paper, Meyer \& Meyer--Hofmeister treat the evaporation
problem in a 1-D vertical way, and obtain semi self--consistent models
of disk evaporation. The goal of this paper is to re--examine this
evaporation mechanism, but without the aim of finding fully consistent
models. In fact, due to the complexity of the gas dynamics in the
evaporation region, it is questionable whether it is possible to model
this process in a consistent way by use of 1-D methods (Abramowicz et
al.  \cite{abrigumlas:1997}). Instead, the goal of this paper is to
study the structure of the boundary layer connecting the ADAF to the
disk, and to study in a qualitative way the mechanism of evaporation,
the conditions under which evaporation is possible, and the extent of
validity of a 1-D vertical treatment of this evaporation process. To
this end, a second order ordinary differential equation (ODE),
containing only the most essential ingredients for the description of
the evaporation process, is derived. This equation is subsequently
reduced to a dimensionless form containing two dimensionless
constants. The resulting dimensionless equation is then solved and a
simple relation between the two dimensionless constants is obtained by
requiring a self--consistency condition to be fulfilled. Finally, by
transforming the dimensionless solutions back to the real problem of
disk evaporation, several important consequences are derived.

\section{Description of the model}
In order to explore the basic physics of evaporation by thermal
conduction, we study the simplest possible situation: a hot plasma
flow on top of an accretion disk, with vertical thermal conduction,
viscous heating, advective cooling and radiative cooling. This
simplified conception of the problem should suffice for the goal of
this paper. In principle one could think of this problem as the
problem of an ordinary ADAF and an SSD co--existing at each radius,
and vertically glued together by a boundary layer. Such a model is
consistent as long as the boundary layer is thin and the evaporation
is not too strong (Dullemond \& Turolla \cite{dulturigum:1997}). Under
these circumstances one can model the ADAF locally as a self--similar
ADAF (Narayan \& Yi \cite{narayanyi:1994}). It is expected that the
qualitative conclusions of the present model also remain valid
somewhat beyond the breakdown of self--similarity. 

In the boundary layer the temperature drops steeply from the ADAF
temperature down to the chromospheric disk temperature. The vertical
scale associated with this gradient is extremely small, much smaller
than the disk thickness. This justifies the use of a 1-D method for at
least the lower parts of the boundary layer. The base of the boundary
layer lies above the chromosphere of the disk, so that an optically
thin treatment of the boundary layer is sufficient. The temperature
gradient constitutes a heat flux pointing downwards towards the
surface of the disk. At relatively high altitudes the ADAF produces an
excess of energy which is transported downwards by the heat flux. In
the lower parts of the boundary layer both radiative cooling and the
upward gas motion absorb the flux. The upward gas motion, acts as a
kind of vertical advective cooling (MMH). It is precisely large
enough to cancel the downward flux, guaranteeing the flux at the base
to be zero. This condition of zero flux at the base determines the
evaporation rate.

As the matter moves upwards from the disk surface, it has the tendency
to shift towards larger radii. This is because, as the temperature
increases, the pressure support against gravity increases. A new
gravitational force balance can only be reached by moving radially
outwards a bit. But this phenomenon has no significant effect on the energy
balance of the boundary layer, and we will ignore it. Also we will
ignore the vertical friction ($t^{z\phi}$).  We do take into account
the decrease of $\Omega$ as matter moves upwards into the corona/ADAF.

The viscosity \, prescription \, is a delicate \, matter in
boundary layers of the type studied here. Friction is assumed to arise
from magneto-turbulence. The length scale associated with this
turbulence is usually chosen ad-hoc, using the famous alpha-viscosity
prescription: $l=\alpha H$. For the disk height $H$ one usually takes
the rough estimate $H=c_s/\Omega_K$, following from vertical pressure
balance.  However, the length scales of magneto turbulence in the
boundary layer are uncertain. If one would take $l=\alpha
c_s/\Omega_K$ then one finds that this length scale exceeds the
boundary layer thickness, $l\gg z$ , close to the lower boundary. This
is inconsistent. In order to avoid having to discuss this highly
uncertain issue, we rather ignore the effects of vertical friction. In
principle the radial friction $t^{r\phi}$ suffers from the same
disease, but its contribution to the energy equation becomes small in
the lower parts of the boundary layer, so this effect can be ignored
here as well.

\section{The equations}\label{sec-equations}
As a lower boundary to the calculational domain I take the height
where the coronal temperature becomes equal to the chromospheric
temperature. The height coordinate $z$ is gauged to zero at this lower
boundary, so that $z=0$ represents the height above the chromosphere. 
As an upper boundary to the calculational domain I take
a height $z_{up}$ obeying roughly $z_{up}\lesssim 0.3 R$ ($R$ being
the radius at which the corona is studied), so that geometric effects
of the flow can safely be ignored. The corona extends well above this
upper boundary, by virtue of the fact that it is assumed to be an
advection dominated flow, for which the thickness $H$ is roughly equal
to the radius $R$. 

The model describes the temperature $T$, the density $\rho$ and the
vertical velocity $v$ as a function of $z$. I presume that the system
is quasi--stationary. The equations for $T$, $\rho$ and $v$ are the
compressible Navier--Stokes equations in the coordinates $R$ (radius),
$z$ (height) and $\phi$ (azimuth), but in this calculation I choose
the corona to be axisymmetric, and self--similar in the radial
direction, thus reducing the problem to a one--dimensional problem in
the coordinate $z$.

The continuity equation is $(\rho v)'=0$ which integrates to
\begin{equation}\label{eq-psi}
\rho v = \frac{1}{2}\Psi
\end{equation}
The radial motion and geometrical terms are neglected. It is assumed
that the radial accretion predominantly takes place at altitude
$z\gtrsim z_{up}$, allowing us to regard the evaporation rate as a
constant of motion in the domain of interest. The integration constant
$\Psi$ is the evaporation rate in units $\gram\,\cm^{-2}\sec^{-1}$.
The factor $1/2$ accounts for the fact that the disk has two sides.
The pressure balance is $(\rho c_s^2)'=0$, where $c_s$ is the
isothermal sound speed. This trivially integrates to
\begin{equation}\label{eq-pi}
\rho c_s^2 = \Pi
\end{equation}
The ram pressure is neglected because the motion is assumed to be
very subsonic. And as in the continuity equation, the geometric terms
are neglected here as well. 

The energy equation consists of five terms: viscous heating $Q_{+}$,
radial advective cooling $Q_{adv}$, optically thin radiative cooling
$Q_{-}$, electro--conductive heat flux $J_c$ and vertical advective
heat flux $J_v$. Viscous dissipation due to the vertical gradient of
the rotational frequency, $d\Omega(z)/dz$, is neglected. 

The dissipation due to the radial gradient in $\Omega$ is given by
\begin{equation}
Q_{+} = \rho \nu \left( R\frac{d\Omega}{dR} \right)^2
\end{equation}
where the kinematic viscosity $\nu$ is given by
\begin{equation}
\nu = \frac{2}{3}\alpha\frac{c_s^2}{\Omega_K}
\end{equation}
By using Eq.~(\ref{eq-pi}) the $Q_{+}$ can be written as
\begin{equation}\label{eq-def-qplus}
Q_{+} = \frac{3}{2}\alpha\omega^2\Omega_K\Pi
\end{equation}
where the dimensionless rotational frequency $\omega$ is defined by
$\Omega=\omega\Omega_K$, with $\omega<1$. For an ADAF, the radial
force balance equation, with omission of the radial kinetic term
(which is very small), is 
\begin{equation}
\Omega^2R^2-\frac{d\log(p)}{d\log(R)}c_s^2 = \Omega^2_KR^2
\end{equation}
We take $p\propto R^{-5/2}$, so that this equation reduces to
\begin{equation}
\omega^{2} = 1 - \frac{5}{2}\frac{c_s^2}{\Omega_K^2R^2}
\end{equation}
This can be substituted in Eq.~(\ref{eq-def-qplus}), for use in the
energy Eq.~(\ref{eq-endifeq}) below. The radial advective cooling
is given by
\begin{equation}\label{eq-qadv-def}
Q_{adv} = \rho v_R \left( \frac{de}{dR} + p \frac{d\rho^{-1}}{dR} \right)
\end{equation}
where $e=c_s^2/(\gamma-1)$ is the thermal energy of the gas.
The radial gas velocity is something that has to be estimated from
the presumed radial structure of the corona. Take $v_R$ to be 
the usual expression
\begin{equation}\label{eq-def-vr}
v_R = -\alpha\frac{c_s^2}{\Omega_KR}
\end{equation}
If one wants to take into account the outwards shift resulting
from the requirement of pressure balance (discussed above),
then one should add the velocity $v_R^{(1)}$ defined as
\begin{equation}
v_R^{(1)} = \frac{5}{2}\frac{R}{\omega^2}\frac{v}{\Omega_K^2R^2}
\frac{d c_s^2}{dz}
\end{equation}
where $v$ is again the vertical velocity. This extra velocity adds a
contribution to the $Q_{adv}$, which is small enough not to influence
the result significantly. For the sake of clarity it is ignored here,
although it is easy to incorporate it. By using equations
(\ref{eq-pi}, \ref{eq-def-vr}), Eq.~(\ref{eq-qadv-def}) becomes
\begin{equation}
Q_{adv}=\alpha\left( \frac{1}{\gamma-1} - \frac{3}{2} \right)
\frac{\Pi}{\Omega_K R^2}c_s^2
\end{equation}
Radiative cooling is denoted with the symbol $Q_{-}$. Several 
cooling mechanisms can can play a role, but the most important
cooling mechanism in the boundary layer is Bremsstrahlung,
$Q_{-}\simeq 5.0\times 10^{20}\rho^2\sqrt{T}$.
The temperature $T$ is related to $c_s^2$ by $kT=\mu m_p c_s^2$,
where $\mu$ is the molecular weight, and $m_p$ is the proton mass. 
The mean particle weights are $\mu=0.59$, $\mu_i=1.23$ and $\mu_e=1.14$.
The Bremsstrahlung cooling function is (using Eq.~(\ref{eq-pi}))
\begin{equation}
Q_{-} = K_1 \Pi^2 c_s^{-3} 
\end{equation}
The constant $K_1$ is defined as
\begin{equation}
\begin{split}
K_1 & \simeq\; 4.2\times 10^{16} \; \cm^6\erg^{-1}\sec^{-4}
\end{split}
\end{equation}
The vertical advective flux is $J_v$, is
\begin{equation}
J_v = \rho v h = \frac{1}{2}\,\frac{\gamma}{\gamma-1} \Psi c_s^2
\end{equation}
where $h$ is the enthalpy, and Eq.~(\ref{eq-psi}) has been used.  The
kinetic energy term has been neglected.  Finally the conductive flux
$J_c \simeq - 9.2\times 10^{-7}\,\mfpeta\,T^{5/2} dT/dz$ (Braginskii
\cite{bragin:1963}, Spitzer \cite{spitzer:1962}) can be written in the
form,
\begin{equation}\label{eq-jc-def}
J_c  = - K_0 c_s^5 \frac{dc_s^2}{dz}
\end{equation}
The conductivity coefficient $K_0$, for conduction along the magnetic
field, is 
\begin{equation}
\begin{split}
K_0 &\simeq\;2.8\times 10^{-35} \; \mfpeta \;\erg\,\sec^6\cm^{-8}
\end{split}
\end{equation}
The Coulomb logarithm is roughly $\ln\Lambda\simeq 20$.  The
coefficient $\mfpeta\le 1$ is put in as a fudge factor to parameterize
the reduction in mean free path length as a result of possible
collective plasma modes and confinement by random magnetic fields. The
microphysics of plasmas in these conditions is insufficiently
understood to allow an estimate of $\mfpeta$ from first principles, so
we must retain it as the main unknown parameter of the model.

The energy equation is now
\begin{equation}\label{eq-endifeq}
\frac{dJ_c}{dz}+\frac{dJ_v}{dz} = Q_{+} - Q_{adv} - Q_{-}
\end{equation}
This is the basic equation of this paper. The density $\rho$ and
the velocity $v$ have all been eliminated in favor of $c_s$, and the
equation has reduced to a single second order diffusion equation. 

The relation between the coronal accretion rate $\dot M_c\equiv 
-4\pi R H\rho v_R$ and the value $\Pi$ is an integral of
the model over the entire vertical height $H$. But within the range of
validity of this model, a good estimate is given by assuming that the
corona can be approximated as a homogeneous flow of height
$H=\sqrt{5/2}\,<\!\!\!c_s\!\!\!>/\Omega_K$ (where $<\!\!\!c_s\!\!\!>$
is the average temperature in the corona). One obtains
\begin{equation}\label{eq-mdot-pi}
% \Pi \simeq \frac{\dot M_c}{4\pi\alpha}\frac{\Omega_K}{R\sigma}
\Pi \simeq \frac{\dot M_c}{2\pi\sqrt{10}}\frac{\Omega_K}{\alpha R\sigma}
\end{equation}
here the symbol $\sigma$ is defined as $\sigma=<\!\!\!c_s\!\!\!>/\Omega_K
R$. It follows from the solution of Eq.~(\ref{eq-en-corona}) below. For
$\gamma=1.5$ and low enough $\dot M_c$ this value is $\sigma\simeq 0.59$.
\section{Dimensionless form}\label{sec-dimless-form}
By defining the variable $y$ as
\begin{equation}
y = \xiadv^{7/2}\left(\frac{c_s}{\Omega_KR}\right)^7
\end{equation}
where $\xiadv$ is
\begin{equation}
\xiadv = \frac{2}{3(\gamma-1)}+\frac{3}{2}
\end{equation}
and a new coordinate $x$ as
\begin{equation}\label{eq-new-x}
x=\sqrt{\frac{21\alpha}{4K_0}}\xiadv^{7/4}
\Pi^{1/2}\Omega_K^{-3}R^{-7/2}\;z
\end{equation}
Eq.~(\ref{eq-endifeq}) becomes
\begin{equation}\label{eq-en-dimles}
-\frac{d^2y}{dx^2}+Dy^{-5/7}\frac{dy}{dx}=1-y^{2/7}-Cy^{-3/7}
\end{equation}
This is the dimensionless form of the main equation of this paper.
From left to right one has the terms representing the divergence of the
conductive flux, the vertical advective cooling, the viscous heating,
the radial advective cooling and finally the radiative cooling. The
symbols $C$ and $D$ are defined as
\begin{eqnarray}
C\!\! &=& \!\!\frac{2K_1\xiadv^{3/2}}{3\alpha}
\frac{\Pi}{\Omega_K^4R^3}\label{eq-def-c-const}\\
D\!\! &=& \!\!\frac{\gamma}{\gamma-1}
\sqrt{\frac{\xiadv^{3/2}}{21\alpha K_0}}
\frac{\Psi}{\sqrt{\Pi}}\frac{1}{\Omega_K^2R^{3/2}}
\end{eqnarray}
Using Eq.~(\ref{eq-mdot-pi}) the constant $C$ can be directly related to
the coronal accretion rate $\dot M_c$ by,
\begin{equation}
C = \frac{K_1\xiadv^{3/2}}{3\pi\sqrt{10}\,\alpha^2\sigma}
\frac{\dot M_c}{\Omega_K^3R^4}
\end{equation}
The values of $\Pi$ and $\Psi$ are,
\begin{eqnarray}
\Pi &=& \frac{3\alpha}{2 K_1\xiadv^{3/2}}\Omega_K^4R^3\,C\\
\Psi &=& \frac{\gamma-1}{\gamma}\sqrt{\frac{63}{2}}\sqrt{\frac{K_0}{K_1}}
\frac{\alpha}{\xiadv^{3/2}}\Omega_K^4R^3\,\sqrt{C}\,D\label{eq-psi-in-cd}
\end{eqnarray}

The solution for the corona at high altitude ($x\gg 1$) must be such
that both fluxes $J_c$ and $J_v$ vanish. The equation is then simply a
balance between viscous heating and advective and radiative cooling,
\begin{equation}\label{eq-en-corona}
1-y^{2/7}-Cy^{-3/7}=0\,.
\end{equation}
This equation is of rank 5 and cannot be solved exactly, but an
very good parabolic approximation (to a few \%) is given by
\begin{equation}
y=\left( \frac{1}{2} \pm \frac{1}{2}\sqrt{1-\frac{125}{6\sqrt{15}}C} \right)^{28/11}
\end{equation}
For $C\leq C_{crit}\equiv 6\sqrt{15}/125$ this equation has two
branches of solutions. The upper branch represents \! the \! ADAF
branch $y_{adaf}$, while the lower branch represents the SLE branch $y_{sle}$
(SLE \cite{sle:1976}). For most of the
solutions discussed in this paper, the corona is found in the
ADAF state, $y(\infty)=y_{adaf}$. At $C=C_{crit}\equiv 6\sqrt{15}/125$
the two branches meet. This value $C_{crit}$ represents the critical
accretion rate for optically thin accretion flows (Abramowicz et
al. \cite{abrchen:1995}),
\begin{equation}\label{eq-mdot-crit}
\dot M_{crit} = \frac{18\pi\sqrt{6}}{25}\,\frac{\alpha^2\sigma}
{K_1\xiadv^{3/2}}\Omega_K^3R^4
\end{equation}
The $\dot M_c$ and \!$C$ relate to each other
as $\dot M_c/\dot M_{crit}=C/C_{crit}$.
\section{Solution without evaporation}\label{sec-no-evap}
As a simple illustration let's first solve the equations without
evaporation, so simply put $D$ to zero. Eq.~(\ref{eq-en-dimles})
becomes
\begin{equation}\label{eq-en-noevap}
-\frac{d^2y}{dx^2}=1-y^{2/7}-Cy^{-3/7}
\end{equation}
At the upper boundary $x=x_{up}$ the thermal conductive flux should
vanishes: $J_c\propto dy/dx=0$.  At the lower boundary $x=0$ the
temperature should vanish, $y=0$. The latter condition is of course
unphysical, since the temperature should equal the chromospheric
temperature rather than zero. But since the chromospheric temperature
is presumed to be very low compared to the coronal temperature, this
approximation is very good (within 1\%).

\begin{figure*}
\epsfxsize=8.2cm \epsfysize=5.5cm
\centerline{\epsfbox{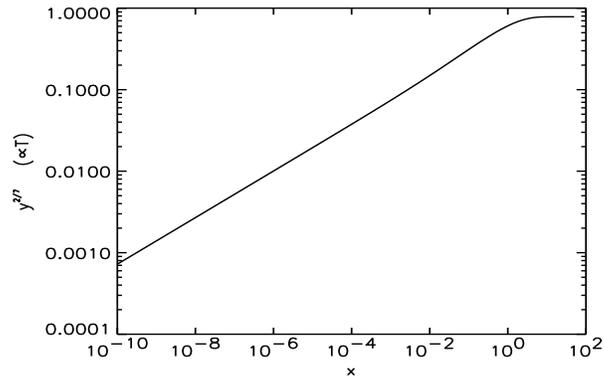}}
\caption{The temperature structure in dimensionless variables. Note
that $y^{2/7}$ is proportional to the temperature. Both axes are
logarithmic to clarify the structure. This solution is for $D=0$ and
$C=0.15$.}\label{fig-sol-no-evap-y}
\end{figure*}

\begin{figure*}
\epsfxsize=8.2cm \epsfysize=5.5cm
\centerline{\epsfbox{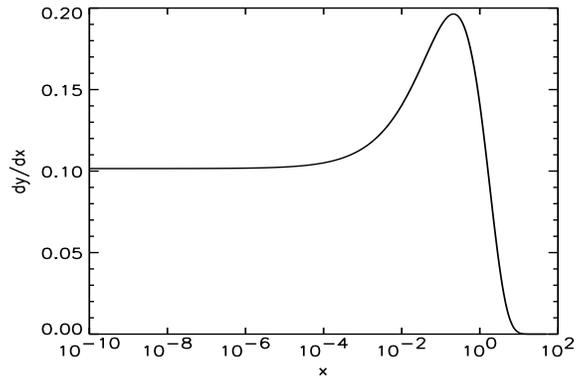}}
\caption{The dimensionless conductive heat flux, plotted linearly. The
solution is the same as for figure \ref{fig-sol-no-evap-y}, for
$D=0$ and $C=0.15$.}
\label{fig-sol-no-evap-dydx}
\end{figure*}

By taking the first integral of Eq.~(\ref{eq-en-noevap}),
one finds the following expression for the dimensionless heat flux 
$j_c\equiv dy/dx\propto J_c$,
\begin{equation}
\begin{split}
j_c^2\equiv\left(\frac{dy}{dx}\right)^2 &= - 2y + \frac{14}{9}y^{9/7}
+ \frac{7}{2}Cy^{4/7}\nonumber\\ & + 2y_{\infty}-\frac{14}{9}y^{9/7}_{\infty}
-\frac{7}{2}Cy^{4/7}_{\infty}\label{eq-first-integral}
\end{split}
\end{equation}
where $y_{\infty}$ is the value of $y(x\rightarrow \infty)$, which is
a solution to Eq.~(\ref{eq-en-corona}).  This equation can be
integrated numerically (using a relaxation scheme, for example), and
the results are plotted in Figs.~\ref{fig-sol-no-evap-y} and
\ref{fig-sol-no-evap-dydx}. The result shows that at $x=x_{up}$ the
corona is in the ADAF state ($y=y_{adaf}$).  Closer to the base of the
corona the temperature drops and the conductive heat flux starts to
grow.  It has a maximum when $y(x)=y_{sle}$, at a certain $x=x_{sle}$. 
For $x>x_{sle}$ there is excess heating and flux is being produced. For
$x<_{sle}$ there is excess cooling and the flux is being absorbed. At
$x=0$ some flux remains unabsorbed. The value of $dy/dx$ at $x=0$ can
be found analytically from Eq.~(\ref{eq-first-integral}),
\begin{equation}
\left.\frac{dy}{dx}\right|_{x=0} = \sqrt{2y_{\infty}-\frac{14}{9}
y^{9/7}_{\infty}-\frac{7}{2}Cy^{4/7}_{\infty}}
\end{equation}
This non-zero flux at the base means that the corona is pumping energy
into the chromosphere. But the chromospheric gas is not able to
radiate it away quickly enough, so it must heat up. This inevitably
leads to evaporation of the upper layers of the chromosphere. In order
to make a more consistent model, the vertical motion should be taken
into account from the start.
\section{Model with evaporation}\label{sec-with-evap}
\begin{figure*}
\epsfxsize=8.2cm \epsfysize=5.5cm
\centerline{\epsfbox{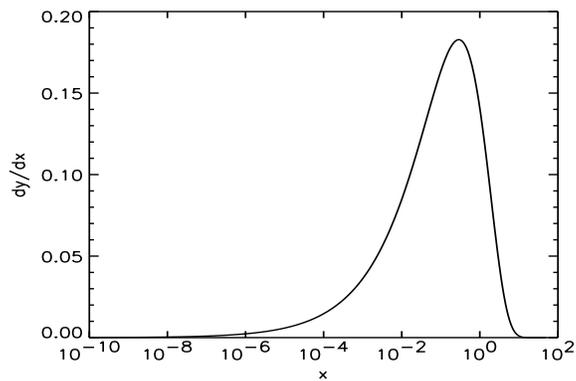}}
\caption{The dimensionless conductive heat flux, plotted linearly. By
requesting the flux at the base to vanish one finds $D=0.2055$ for
$C=0.15$.}
\label{fig-sol-yes-evap-dydx}
\end{figure*}

An upwards motion of the gas constitutes an additional vertical
advective cooling (MMH).  By allowing the constant $D$ to be
non-zero, this cooling takes effect. The value of $D$ (being a
constant over the entire domain) is determined by adding an additional
boundary condition to the system,
\begin{equation}
\left.\frac{dy}{dx}\right|_{x=0} = 0\,.
\end{equation}
This yields both a solution for $y(x)$ and a value for $D$. The
evaporation rate is therefore found as an eigenvalue of the system,
similar to the case of interstellar cloud evaporation studied by Cowie
\& McKee (\cite{cowmckee:1977}) and McKee \& Cowie
(\cite{mckeecow:1977}). The solution for the flux $dy/dx$ is plotted
in Fig.~\ref{fig-sol-yes-evap-dydx}. One can identify three regions
in order of decreasing $x$,
\begin{enumerate}
\item\label{region-up} For $x_{eq}\lesssim x<x_{up}$ the corona is nearly in 
local heating/cooling balance. It is a solution of Eq.~(\ref{eq-en-corona}).
\item\label{region-finc} For $x_{max}<x\lesssim x_{eq}$ the heat flux rises
from almost zero to a maximum at $x_{max}$, as a result of an excess in
local heat production.
\item\label{region-fdec} For $0< x<x_{max}$ the heat flux decreases
again to zero as a result of both radiative cooling and evaporation.
For this region the solution is approximately a power-law, $T\propto
y^{2/7}\propto x^{2/5}$ and the flux goes as $J_c\propto -
dy/dx\propto x^{2/5}$.
\end{enumerate}

\begin{figure*}
\epsfxsize=8.2cm \epsfysize=5.5cm
\centerline{\epsfbox{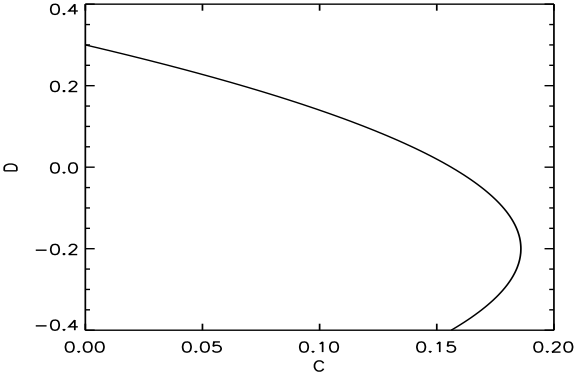}}
\caption{The relation between the dimensionless evaporation rate $D$
and the coronal accretion rate in the form of $C$.}
\label{fig-sol-c-d}
\end{figure*}

The value of $D$, following from these models, depends on the value of
$C$, or in other words on the coronal accretion rate $\dot M_c$.  This
functional dependence of $D$ on $C$ is plotted in Fig.~\ref{fig-sol-c-d}. The curve is closely fitted (to within a few \%) by
the expression
\begin{equation}
C = \frac{6}{125}\sqrt{15}\left[1-4\left(D+\frac{1}{5}\right)^{2}\right]
\end{equation}
Note that, although the fit is very good, it is not exact.  This
expression then gives the evaporation rate $D$ as a function of
$C$. There are two branches. The upper branch has evaporation for
small $C$ (small $\dot M_c$) and condensation for large $C$ (large
$\dot M_c$). For this upper branch the corona (region \ref{region-up})
is in the ADAF mode. The lower branch only has condensation, and the
corona is in the SLE mode.  Since SLE flows are known to be thermally
unstable, this lower branch is not likely to represent any physical
situation. The dimensionless evaporation rate for the upper branch is then
\begin{equation}\label{eq-d-in-c}
D = \frac{3}{10} \left[ - \frac{2}{3}
+\frac{5}{3}\sqrt{1-\frac{125}{6\sqrt{15}}C} \right]
\end{equation}
The final formula for the evaporation rate is then (using
Eqs.~(\ref{eq-psi-in-cd}) and (\ref{eq-mdot-pi})),
\begin{equation}\label{eq-psi-in-mdot}
\Psi=\frac{3}{10}\sqrt{\frac{21}{2\pi\sqrt{10}}}\frac{\gamma-1}{\gamma}
\sqrt{K_0}\frac{\Omega_K^{5/2}R}{\xiadv^{3/4}\sqrt{\sigma}}\,f(C/C_{crit})
\,\sqrt{\dot M_c}
\end{equation}
where $f(C/C_{crit})\equiv (10/3)\,D(C)$ is the saturation function, which is
unity for $C\ll C_{crit}$, i.e. for $\dot M_c\ll \dot M_{crit}$.

There is a distinct value of $C$ for which no evaporation nor condensation
takes place,
\begin{equation}
C_{0}=\frac{21}{25}C_{crit}
\end{equation}
which corresponds to $\dot M_c=(21/25)\dot M_{crit}$. For a more
dilute corona there will be evaporation, while for a denser corona
there will be condensation. It is therefore tempting to conclude that
the corona will always tend towards saturation at $\dot M_c =
(21/25)\dot M_{crit}$. But there is a caveat here, which will be
discussed below.

Before concluding this section it is interesting to compare the
present model with the model of MMH. In the model of MMH part of
the evaporated material leaves the system through a transsonic wind,
while the other part accretes radially onto the central object. By
making a plausible assumption for the radial derivatives of the
density and velocity, the evaporation problem is reduced to a 1--D
vertical problem. The temperature in the wind is fixed by the
conditions at the sonic point, while the pressure at the base, and
thereby the evaporation rate, is determined by the balance between
wind mass loss and conductive evaporation rate. In this way a
well--determined expression for the evaporation rate as a function of
radius can be given.

In the present model, a transsonic wind is not considered, although it
is not ruled out. The temperature at high $z$ is fixed by the
condition that viscous heating is balanced by radial advective cooling
and radiative cooling. The evaporation rate is kept a function of the
pressure, which is related to the radial accretion rate in the
corona. Instead of making any assumption of the radial derivative of
density, we have derived a relation between the evaporation rate and
the accretion rate in the corona. In the next section we will relate
these to the the conductive scale height and to a measure for the
effectiveness of evaporation.
\section{Efficiency and conductive scale height}\label{sec-efficiency}
The corona can only reach saturation \!($\dot M_c\!=\!(21/25)\dot M_{crit}$)
when the evaporation is efficient enough. The evaporation rate has to
compete with radial `mass loss', i.e. the flow of coronal matter
towards the central object. If a corona cannot evaporate the disk
efficiently enough, the radial coronal flow will deplete the corona,
until a low enough coronal density is reached for evaporation to
compete with radial inflow.

To investigate this one should compare the evaporation rate $\Psi$
with the coronal accretion rate $\dot M_c$. Coronal mass conservation
in a stationary situation is given by
\begin{equation}
\frac{d\dot M_c}{dR} = -2\pi R \Psi
\end{equation}
Define an effectiveness index $\myxi$ as $\myxi\equiv
-d\log\dot M_c/d\log R$. One has
\begin{equation}\label{eq-xi-def}
\myxi = 2\pi R^2 \frac{\Psi}{\dot M_c}
\end{equation}
The dimensionless number $\myxi$ gives the ratio between the
evaporation rate and the coronal depletion rate due to radial
accretion. One can also think of it as the ratio between the time
scales of vertical motion and radial motion. For $\myxi<0.5$ the
evaporation rate is weak and for $\myxi>0.5$ it is strong. It should
be kept in mind that for $\myxi\ge 0.5$ there exist no self--similar
ADAF coronae (Dullemond \& Turolla \cite{dulturigum:1997}), so that the
evaporation models for $\myxi>0.5$ are not self--consistent with
respect to angular--momentum conservation. By using
Eqs.(\ref{eq-xi-def}, \ref{eq-psi-in-mdot}, \ref{eq-mdot-crit}) one
obtains
\begin{equation}
\begin{split}
\myxi&=\frac{3}{2}\sqrt{\frac{7}{6\sqrt{15}}}\frac{c\sqrt{K_0K_1}}
{\alpha\sigma}\frac{\gamma-1}{\gamma}\sqrt{\frac{R_g}{R}}\,
\frac{f(\dot M_c/\dot M_{crit})}{\sqrt{\dot M_c/\dot M_{crit}}}\\
&\equiv\bar\myxi\,f(\dot M_c/\dot M_{crit})
=15.1\;\frac{\sqrt{\mfpeta}}{\alpha}\sqrt{\frac{R_g}{R}}\,
\frac{f(\dot M_c/\dot M_{crit})}{\sqrt{\dot M_c/\dot M_{crit}}}
\end{split}
\end{equation}
where $R_g=2GM/c^2$. In the last step $\gamma=1.5$.

Closely related to the evaporation efficiency is the conductive scale
height $z_c$. This is the height below which the coronal energy
content is significantly drained by thermal conduction.  By using
Eqs.(\ref{eq-new-x}, \ref{eq-mdot-crit}) one can find the 
dimensionless conductive scale height $\zeta\equiv z_c/H$,
\begin{equation}
\begin{split}
\zeta&=\frac{10}{21}\sqrt{\frac{7}{3\sqrt{15}}}
\frac{c\sqrt{K_0K_1}}{\alpha\sigma\xiadv}
\sqrt{\frac{R_g}{R}}\frac{x_{c}}{\sqrt{\dot M_c/\dot M_{crit}}}\\
&=7.2\;\frac{\sqrt{\mfpeta}}{\alpha}
\sqrt{\frac{R_g}{R}}\frac{x_{c}}{\sqrt{\dot M_c/\dot M_{crit}}}
\end{split}
\end{equation}
The constant $x_{c}$ is the conductive scale height in the
dimensionless $x$-coordinate, introduced in section
\ref{sec-with-evap}. As one can infer from the figures, the flux is
roughly 20\% of its maximum value at $x\simeq 4$, so take the constant
$x_{c}=4$. For $\zeta\ll 1$ the thermal conduction only affects
the lower layers of the corona, while for $\zeta\gtrsim 1$ the conduction
affects the entire corona. In fact, if $\zeta\gtrsim 1$ one should expect
radial thermal conduction to play an important role. 

If $\dot M_c\ll \dot M_{crit}$, the two dimensionless numbers
$\myxi=\bar\myxi$ and $\zeta$ are roughly equal. For $\gamma=1.5$ and
$x_{c}\simeq 4$ one finds
\begin{equation}
\zeta\simeq 1.9 \myxi
\end{equation}
One sees that when the evaporation efficiency tends to become strong
($\myxi\gtrsim 0.5$), the conductive scale height tends to exceed the
size of the system $\zeta\gtrsim 1$. This shows that for strong
evaporation the conduction will dominate the entire corona from top to
bottom, and radial thermal conduction will be an important effect.
So, it is to be expected that for strong evaporation the problem
becomes essentially 2-D. A similar conclusion has already been drawn
from angular momentum conservation considerations (Abramowicz et
al.~\cite{abrigumlas:1997}, Dullemond \& Turolla
\cite{dulturigum:1997}), but in that case a global radial 1-D solution
with locally computed evaporation could still not be convincingly
excluded. The present conclusion is more dramatic, since it is
unlikely that a very non--linear phenomenon like thermal conduction
lends itself for dimensional splitting.

Now consider the case when the corona becomes saturated, $\dot
M_c\rightarrow(21/25)\dot M_{crit}$. Since $\dot M_{crit}\propto
R^{-1/2}$, this saturation can be only achieved when $\myxi = \bar
\myxi\, f(\dot M_c/\dot M_{crit}) = 0.5$, which implies $\bar\myxi\ge
0.5$ (since $f$ obeys $f(\dot M_c/\dot M_{crit})\le 1$). This can be 
achieved for $R\lesssim 10^3
(\mfpeta/\alpha^{2})\,R_G$, but for most radii
this means that $\bar\myxi\gg 1$ and therefore $\zeta\gg 1$. So one
should conclude that coronal saturation is necessarily accompanied by
strong radial conduction. Under those circumstances the present model
breaks down, and fully 2-D models should be used instead.
\section{Conclusion}
The model presented in this paper described qualitatively the
mechanism of disk surface evaporation by thermal conduction. It
applies to accretion disc systems around black holes, neutron stars
and white dwarfs. Although the model cannot be applied to the case of
evaporation very close to a black hole (because of the relativistic
velocities of the electrons, and the 2-temperature nature of the
plasma) the qualitative picture sketched by this model is applicable
to that case as well.

In order for a disk to evaporate completely, the evaporation should be
efficient. In a relatively short interval in radius, the complete disk
should vanish, and the advection dominated corona should be produced.
The corona provides the energy for achieving this, and therefore the
corona itself determines how strong the evaporation will be. In an
equilibrium situation the evaporation rate should balance the rate of
radial inflow of coronal material. If most of the evaporation takes
place in a relatively small range in radius, this means that the
vertical velocity of the evaporating gas is in the order of the radial
accretion velocity of the corona. The model described in this paper
shows that in order to achieve this strong evaporation rate, the
conductive scale height should be equal or larger than the
radius. Consequently, radial thermal conduction will enter the
problem. The heat required for evaporation does not come from the
upper layers of the corona anymore, but is instead produced closer to
the central object and radially transported to the evaporation region.
There the flux will be directed down towards the disk surface and
evaporate the disk. 

In order to model this complete disk evaporation with radial
conduction, one should take the radial dimension of the problem into
account. One could think of using the usual dimensional splitting
procedure to solve the evaporation vertically (using a local 'heating'
term to account for the divergence in radial heat flux), and the
coronal dynamics and radial thermal conduction radially. Unfortunately
this is procedure is highly questionable because of the very
non-linear nature of the formula for conductive heat flux. It seems
therefore that one should conclude that disk evaporation by thermal
conduction is an essentially 2-D process.

These conclusions do not depend on the micro physical assumptions for
evaporation, like the value of the conductivity suppression factor
$\mfpeta$. The conclusions here can be traced back to energy budget
considerations.

If electron thermal conduction is indeed responsible for the
evaporation of a Shakura--Sunyaev disk, then there are also important
consequences for the spectral modeling of ADAFs.  If Comptonized
soft-photons from the SSD constitute an important part of the spectrum,
then this part emerges from the ADAF region that is strongly affected
by radial thermal conduction. It is therefore questionable whether a
consistent 2-phase (SSD+ADAF) spectral model can be built without 2-D
radiative--hydrodynamic simulations.

\begin{acknowledgements}
I am much indebted to I.V.~Igumenshchev for interesting discussions on
the topic of disk evaporation, and on the interpretations of the
results of this model. I received valuable help with the differential
equations by V.~Icke and R.~Turolla. I also thank A.~Helmi, C.~v.~Duin
and Y.~Simis.
\end{acknowledgements}

\newcommand{\etalchar}[1]{$^{#1}$}

%\pagebreak[4]

\end{document}